\documentclass[superscriptaddress,nofootinbib,twocolumn]{revtex4-2}
\usepackage{amssymb}
\usepackage{amsmath}
\usepackage{graphicx}
\usepackage{mathdots}
\usepackage{slashed}
\usepackage{verbatim}
\usepackage{xcolor}

\newcommand{\be}{\begin{equation}}
\newcommand{\ee}{\end{equation}}
\newcommand{\bea}{\begin{eqnarray}}
\newcommand{\eea}{\end{eqnarray}}

\begin{document}

\title{RG Dynamics of Irrelevant Fermion Operators and the
 Drag‑Coupling Mechanism}

\author{J. Aguilera-Damia} \affiliation{ Departament de Física Quàntica i Astrofísica and Institut de Ciències del Cosmos, Universitat de Barcelona, Martí i Franquès 1, 08028 Barcelona, Spain } \author{Diego Rodriguez-Gomez} \affiliation{ Department of Physics, Universidad de Oviedo, C/ Federico García Lorca 18, 33007 Oviedo, Spain } \affiliation{ Instituto Universitario de Ciencias y Tecnologías Espaciales de Asturias (ICTEA), C/ de la Independencia 13, 33004 Oviedo, Spain } \author{J. G. Russo} \affiliation{ Departament de Física Quàntica i Astrofísica and Institut de Ciències del Cosmos, Universitat de Barcelona, Martí i Franquès 1, 08028 Barcelona, Spain } \affiliation{ Institució Catalana de Recerca i Estudis Avançats (ICREA), Pg. Lluís Companys 23, 08010 Barcelona, Spain }

\begin{abstract}
We study the renormalization-group flow of higher-dimensional fermionic interactions
$(\psi^\dagger \psi)^{2n}$ in the presence of a Fermi surface. We show that the growth
of the BCS four-fermion coupling induces a drag mechanism whereby higher-order
fermionic couplings are driven to strong coupling in the infrared. 
We derive the corresponding beta functions and show that the drag effect
applies generically to the whole tower of fermionic operators. Remarkably, although
all couplings are driven toward the same strong-coupling scale, the renormalization-group
flow preserves a hierarchy in which higher-dimensional operators remain parametrically
suppressed relative to the BCS interaction. We then investigate this mechanism in a
$2+1$-dimensional non-Fermi liquid coupled to a critical boson. While higher-order
fermionic interactions are similarly enhanced along the flow, we show that they do not
destabilize the IR stable non-Fermi-liquid fixed point, present for sufficiently large $N$.
We briefly discuss possible implications for multicomponent superconductors and other
strongly correlated metallic systems.

\end{abstract}

\maketitle


\section{Introduction}

The Renormalization Group (RG) flow plays a central role in modern theoretical physics, as it keeps track of the relevant degrees of freedom while the scale at which a system is probed is varied. 
Operators are conventionally classified as relevant, marginal, or irrelevant under the renormalization group flow. Irrelevant operators are expected to have negligible influence on the infrared behaviour of the theory. This expectation, however, is not universal, and notable exceptions are known.
 A paradigmatic example is BCS theory of superconductivity, where it appears, at first sight, that a phonon‑mediated four‑Fermi interaction—being power‑counting irrelevant—should be unable to destabilize the Fermi gas and drive it into a non‑trivial superconducting state. Nevertheless, the presence of a Fermi surface profoundly modifies the RG flow,  transmuting an otherwise irrelevant operator into a (marginally) relevant one \cite{Polchinski,Shankar}. 
Very recently, a similarly striking manifestation of this phenomenon was identified in \cite{Rodriguez-Gomez:2025hyy}, where it was argued that, in the presence of a four‑Fermi interaction, higher‑dimensional fermionic operators such as 
 $(\psi^\dagger \psi)^{4}$
-- which are naively far more irrelevant -- can also be dragged toward strong coupling at low energies, potentially leading to observable effects. The purpose of this note is to place this observation on firmer footing by analysing in detail the RG flow of these higher‑dimensional fermionic operators.

In the case of the Fermi gas, the standard BCS four‑Fermi interaction becomes (marginally) relevant and drives the system into a superconducting phase at low energies. Higher‑dimensional fermionic couplings are likewise dragged toward strong coupling and can modify the properties of the resulting low‑energy phase. 

One may then ask whether this dragging of higher‑dimensional operators could itself destabilize 
-- or at least significantly alter -- an otherwise stable fixed point. Motivated by this question, we turn to the quantum‑critical metallic system with non‑Fermi‑liquid dynamics in two spatial dimensions studied in \cite{Damia:2019bdx,Damia:2020yiu,Damia:2020bur} (other studies  include \cite{Metlitski2010,Lee2009, Mross2010}). The model consists of an 
$N$-component fermion coupled to a critical boson. This class of strongly coupled systems typically flows to interesting quantum critical points \cite{Hertz1976, Altshuler1994, Nayak1994}, whose low energy excitations are not captured by standard Landau quasiparticles \cite{Schofield1999}. Moreover, these non-Fermi liquid fixed points are believed to underlie interesting phase transitions, such as the Ising-nematic \cite{Shibauchi2013, Kuo2016}, and may also be relevant in the context of High Critical Temperature Superconductivity \cite{Kivelson-Zaanen-review}.  In the presence of a four‑Fermi interaction, the system exhibits nontrivial RG trajectories, including a stable interacting fixed point. 
Within this richer framework, we examine how higher‑order fermionic couplings are dragged along the flow and influenced by the dynamics of the lower‑dimensional operators.

\section{Multiflavor BCS theory extended by a 
$(\psi^\dagger\psi)^4$ coupling}\label{sec:Fermi}

We consider a non-relativistic 2+1 dimensional theory with $N$ fermions   transforming in the fundamental representation of a $U(N)$ flavour symmetry, coupled to a chemical potential.
The most general Lagrangian invariant under $U(N)$ symmetry has the form
\begin{equation}\label{eq: gen Lag}
L= \psi^{\dagger}_i\, \left( \partial_t+\frac{\vec\partial^2}{2m}+\mu_F\right)\psi^i-\sum_{n=1}^\infty 
\frac{g_{4n}}{(2n)! N^{2n-1}}\,\left(\psi^{\dagger}_i\psi^i\right)^{2n} .
\end{equation}
The possible number of terms is limited due to the Grassmann character of the fermion variables.
In the case $N=2$ there is a unique $\psi^4$
interaction and the Lagrangian describes BCS theory
(the two species of fermions organize into a single two-component spinor). When $N=4$ (i.e. two spinors), there is a new $\psi^8$
interaction of the form $\psi^1\psi^2\psi^3\psi^4\psi_1^\dagger\psi_2^\dagger\psi_3^\dagger\psi_4^\dagger$.
Possible higher interactions appear for larger $N$.
In the absence of a Fermi surface (FS), all these operators are irrelevant and do not have a significant impact
in the IR physics. 
In particular, the operators 
$$
{\cal O}_4=\int d^3 x \left(\psi^{\dagger}_i\psi^i\right)^2\ ,\qquad 
{\cal O}_8=\int d^3 x \left(\psi^{\dagger}_i\psi^i\right)^4\ ,
$$ have dimension 1 and 5, respectively.
The presence of a FS leads to a dynamics that modifies the
effective dimensions of the operators.
 The leading contribution to the quasiparticle energy then comes from the momentum transverse to the FS,
and this naturally implies the RG scaling \cite{Polchinski,Shankar}
\be\label{eq: FS scaling}
t\to s^{-1} t \quad , \quad k_\perp\to s k_\perp \quad , \quad  k_\parallel \to  k_\parallel\ ,
\ee
towards the FS as $s\to 0$. 
This describes the fact that
low energy excitations correspond to  electrons 
moving near the Fermi surface, which implies
$|k_\perp|\ll |k_\parallel|$.

It is convenient to introduce the Fourier components
\be
\tilde \psi(t,k_\perp,k_\parallel)=\int dk_\perp dk_\parallel\ \psi(t,\vec x) \, e^{i\vec k\cdot \vec x}\ .
\ee
Scale invariance of the kinetic term now implies that $\tilde \psi$ has dimension $\frac12$, i.e. it scales as $\tilde \psi \to s^{-1/2} \tilde\psi$. Now we can make use of the above prescription to evaluate the RG scaling dimension of the various interactions. A general ($U(N)$ symmetric) $4n$-fermi interaction schematically reads (for simplicity in the notation, for the remainder of this work we drop the tildes from $\tilde \psi$'s)
\be\label{eq: general vertex}
V_{4n}=\int  d\mu_n \, \psi(\vec k_1)\ldots \psi(\vec k_{2n}) \psi^\dagger(\vec k'_1)\ldots \psi^\dagger(\vec k'_{2n}) \delta_{\rm mc}\, ,
\ee 
\be
d\mu_n\equiv dt \prod_{i}^{2n} d^2 k_i d^2 k'_i\ ,
\ee
where $\delta_{\rm mc}$ denotes the two-dimensional delta function imposing the total momentum conservation $\sum_i\vec k_i=\sum_i \vec k'_i$. 

Consider a rotationally invariant (circular) FS. 
It is convenient to parametrize the fermion momentum as $\vec{k}=(k_F+k_\perp)\vec{n}$ with $\vec{n}=(\cos\theta,\sin\theta)$ and $k_\perp\in [-\Lambda,\Lambda]$ where $\Lambda\ll k_F$. In that regime, we can expand $\epsilon_k=v_{0}\,k_\perp$, where $v_0=k_F/m$ denotes the (bare) Fermi velocity.

Assuming that $\delta_{\rm mc}\to s^{-[\delta]} \delta_{\rm mc}$ under the rescaling \eqref{eq: FS scaling} one finds
\be\label{eq: 2n scaling}
V_{4n} \, \to \, s^{2n-1-[\delta]} V_{4n}\ .
\ee      
For general kinematics, one can neglect the orthogonal components $k^\perp_i$ in the argument of the $\delta$
function. Thus the argument only contain parallel components that have trivial scaling. In this case
$[\delta]=0$.
This implies that
\be
[V_{4n}]=2n-1\ .
\ee
In particular, $[V_{4}]=1$ and $[V_{8}]=3$. The latter
dimension is already less than the canonical dimension 5 of the operator when there is no Fermi surface.
However, both operators are still irrelevant.

Nevertheless, there are special configurations for which the argument of $\delta_{\rm mc}$ is proportional to the transverse components $k^{\perp}_i$ only. In such a  case, $[\delta]=1$ and the $4n$-fermion operator has dimension $2n-2$. The argument of the momentum-conservation delta function in \eqref{eq: general vertex} is
\be
\sum_i^{2n}\vec{k}_i-\sum_i^{2n}\vec{k}'_i= k_F\left(\sum_i^{2n}\vec{n}_i-\sum_i^{2n}\vec{n}'_i\right) + \mathcal{O}(k_{\perp})\ ,
\ee
where the last piece comprises the terms proportional to $k_\perp$. A non-trivial scaling of the delta function is therefore attained for kinematical configurations satisfying
\be\label{eq: scaling condition}
\sum_i^{2n}\vec{n}_i=\sum_i^{2n}\vec{n}'_i\ .
\ee
A general solution requires fixing two of the $4n$ angular variables.
For $n=1$, according to \eqref{eq: 2n scaling}, one is left with a marginal four-fermion interaction when the above condition is met.  A particularly important solution is attained for $\vec n_1+\vec n_2=\vec {n}'_1+\vec n'_2=0$ leading to the marginal BCS 4-fermion coupling\footnote{The other independent marginal channel corresponds to forward scattering $\vec{n}_1=\vec{n}'_1$. This channel does not lead to any instability and we will not consider it further.}. Furthermore, logarithmic quantum corrections trigger a relevant running in this interaction channel, eventually leading to a pairing instability.

Similarly, the 8-fermion operator ($n=2$) will have dimension $\Delta_8=2$ in the kinematical corners where
the momenta satisfy the condition \eqref{eq: scaling condition}. In addition, as we will show below, there is a particular kinematic corner that leads to a logarithmic divergence in one-loop diagrams, namely
\be\label{eq: 8 fermi BCS}
\vec{n}_1=-\vec{n}_2 \,\, , \,\, \vec{n}_3+\vec{n}_4=\sum_i^{4}\vec{n}'_i\ .
\ee
 Along this channel, the 8-fermion coupling scales as $s^2$. These kinematics are particularly significative in the context of this article, as they enable the ``drag" effect induced by the 4-fermi BCS coupling. We will henceforth restrict our attention to this particular interaction channel.   

\begin{figure}[h!]
 \centering
 \includegraphics[scale=.25]{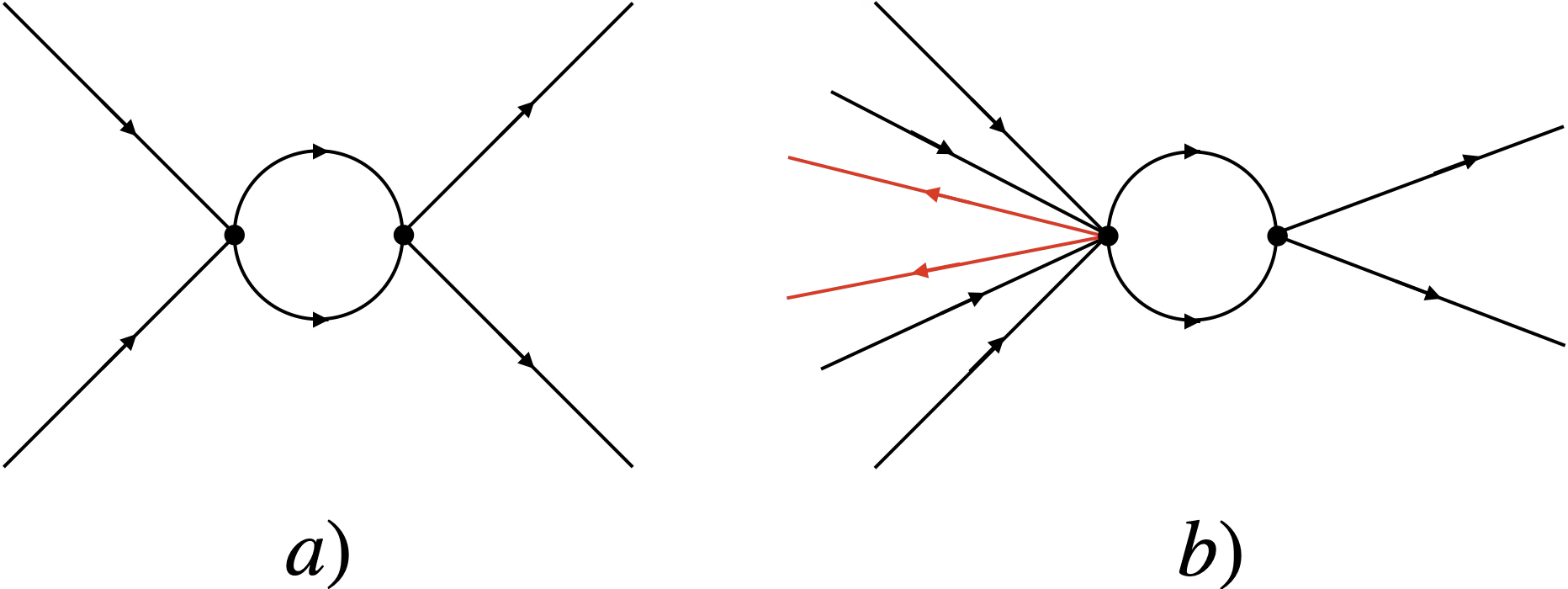}
 \caption{One-loop contributions to $a)$ the 4-fermion BCS channel, $b)$ 8-fermion function (holes are depicted in red).}
 \label{Diagrams}
 \end{figure}

In order to analyse the effective running of these couplings, we introduce the sliding scale $\mu$ such that the flow to the IR is accounted for by $\mu\to 0$. The leading Feynman diagrams at one-loop contributing to the running of the 4- and 8-fermion couplings are shown in Fig.~\ref{Diagrams}.

In both diagrams the running is induced by the same fermion loop. The calculation of the fermion loop integral is a well known result \cite{Polchinski} and we briefly reproduce it here. Given the bare fermionic propagator expanded around the Fermi surface
\be
G_F(\omega,\vec{k})=\frac{1}{i\omega-v_0 k_\perp}\,,
\ee
the fermion loop for vanishing total incoming momenta reads
\begin{align}\label{eq: fermion bubble}
-\frac{g_4^2}{N^2}&\int \frac{d\omega}{(2\pi)}\int \frac{d^2\vec{k}}{(2\pi)^2}\,G_F(-\omega,-\vec{k})\,G_F(\omega,\vec{k}) \\
&=- \frac{g^2_4 k_F}{N^2}\int \frac{d\omega dk_\perp d\theta}{(2\pi)^3} \frac{1}{-i\omega +v_0 k_\perp}\frac{1}{i\omega +v_0 k_\perp}  \nonumber \\
&= -\frac{g_4^2 k_F}{2\pi v_0 N^2} \log\left(\frac{\Lambda}{\mu}\right)\nonumber
\,\, .
\end{align}

Using this result, one obtains that the one-loop beta function for the BCS 4-fermion coupling is
\be\label{eq: 4-fermi beta}
\mu\partial_\mu \tilde g_4= -\frac{1}{2\pi v N} \tilde g_4^2\ ,
\ee
where $g_4=k_F \tilde g_4$ and $\tilde g_4$ is dimensionless. Note that we have replaced the bare Fermi velocity by the physical one, as it does not get renormalized within this model, as opposed to the non-Fermi liquid model that we will consider in the next section.

As for the eight‑fermion interaction, a qualitative discussion of its expected running was presented in \cite{Rodriguez-Gomez:2025hyy}. We now provide the explicit calculation of the one‑loop contribution to the corresponding beta function, including the precise coefficients.
The relevant diagram is that of panel $b)$ in Fig.~\ref{Diagrams}. Like in BCS theory, a logarithmic divergence appears for vanishing incoming/outgoing total momenta which, in particular, is attained at low energies within the channel \eqref{eq: 8 fermi BCS}.

Making use of \eqref{eq: fermion bubble} and taking into account the combinatorial factor equal to $2\times 4!$,
we obtain the following beta function
\be\label{eq: beta 8}
\mu\partial_\mu\tilde g_8 = \Delta_8 \tilde g_8-\frac{1}{2\pi v N} \tilde g_4\tilde g_8+\mathcal{O}(\tilde g_8^2)
\ee
where we have defined the dimensionless coupling 
$$
\tilde g_8=k_F^{5-\Delta_8}\mu^{\Delta_8}g_8\ ,
$$
where $\Delta_8$ represents the tree-level scaling dimension, and the powers of $k_F$ are again introduced  to match units. Generically $\Delta_8$ equal to 3, but it is reduced to $\Delta_8=2$ for the special operators containing fermions with momenta satisfying \eqref{eq: 8 fermi BCS}.

The solutions to this coupled system of differential equations take the form
\be\label{eq: effective lambda4}
\tilde g_4(\mu)=\frac{\tilde g_{4,0}}{1-\frac{\tilde g_{4,0}}{2\pi  v N} \log\frac{\Lambda}{\mu}}\ ,
\ee
\be\label{eq: effective lambda8}
\tilde g_8(\mu)=\left(\frac{\mu}{\Lambda}\right)^{\Delta_8}\frac{\tilde g_{8,0}}{1-\frac{\tilde g_{4,0}}{2\pi  v N}\log\frac{\Lambda}{\mu}}\ .
\ee
Remarkably,  the growth of the BCS coupling $\tilde g_4$ towards the IR  drives the  8-fermion coupling to large values. This occurs because of the mixed term in \eqref{eq: beta 8}, which drags the $\tilde g_8$ to strong coupling when $\tilde g_4$ becomes strong.
As a result, the running coupling $\tilde g_8(\mu)$ has a pole at the same BCS strong coupling scale, overtaking the classical power-like behaviour $\mu^{\Delta_8}$.

Note that 
\be
\frac{\tilde g_8(\mu)}{\tilde g_4(\mu)}= {\rm const.}\ \mu^{\Delta_8}\to 0\ ,\qquad {\rm as \ }\mu\to 0\ .
\ee
This simple ratio is a striking consequence of
the combinatorial factor being equal to $2\times 4!$
in the diagram of Fig. \ref{Diagrams}(b).
If the factor was $2\times 4!\times k$, for some rational number $k$,
the solution for $\tilde g_8(\mu)$ would have a power $k$ in the denominator, giving

\be
\frac{\tilde g_8(\mu)}{\tilde g_4(\mu)}= {\rm const.}\ \frac{\mu^2}{\left(1-\frac{\tilde g_{4,0}}{2\pi  v N}\log\frac{\Lambda}{\mu}\right)^{k-1}}\ .
\ee
For  $k> 1$, this expression diverges as 
$\mu$ is lowered, eventually reaching the singular point in the denominator. This would imply the unexpected result that the coupling  $\tilde g_8(\mu)$
 becomes relatively more significant than 
 $\tilde g_4(\mu) $ at low energies.
This would be a strong departure from BCS behaviour and, more generally, from the expectations of effective field theory.
It is therefore remarkable that the diagram in Fig.\ref{Diagrams}(b) yields precisely the coefficient needed for the cancellation to occur, ultimately producing such a simple final ratio.

Let us now consider the interaction $g_{4n}\,(\psi^{\dagger}_i\psi^i)^{2n}$ in \eqref{eq: gen Lag}. The leading one-loop  contribution to the beta function  originates from the diagram analogous to \ref{Diagrams}(b). Again we consider the kinematic configuration $\vec n_1=-\vec n_2$.
The combinatoric factor is now $2\times (2n)!$.
This leads to a beta function 
\be\label{eq: beta 4n}
\mu\partial_\mu\tilde g_{4n
} = \Delta_{4n} \tilde g_{4n}-\frac{1}{2\pi v N} \tilde g_4\tilde g_{4n}+\mathcal{O}(\tilde g_{4n}^2)\ ,
\ee
with solution
\be\label{lambda4n}
\tilde g_{4n}(\mu)=\left(\frac{\mu}{\Lambda}\right)^{\Delta_{4n}}\frac{\tilde g_{4n,0}}{1-\frac{\tilde g_{4,0}}{2\pi  v N}\log\frac{\Lambda}{\mu}}\ .
\ee
Here $\Delta_{4n}=2n-1$ generically and $\Delta_{4n}=2n-2$
for the operators considered here with momentum satisfying the general condition \eqref{eq: scaling condition}. 
As mentioned before, within these kinematics we focus on the region for which $\vec n_1=-\vec n_2$ which gets logarithmically enhanced at one-loop.
Thus, we   again find the strikingly simple ratio
\be
\frac{\tilde g_{4n}(\mu)}{\tilde g_{4}
(\mu)}= {\rm const.}\ \mu^{\Delta_{4n}}\to 0\ ,\qquad {\rm as \ }\mu\to 0\ ,
\ee
where the potentially problematic pole has cancelled out.

\medskip

In conclusion, the renormalization group causes the drag effect to appear in all couplings 
$\tilde g_{4n}$, driving them to strong coupling at low energies, but with a hierarchy determined by the effective dimension of each operator. This picture complements the mean field analysis carried out in \cite{Rodriguez-Gomez:2025hyy} by numerically solving the corresponding Schwinger-Dyson equations derived from a modified BCS theory. 

The significance of the  8-fermion coupling has been discussed in
\cite{Rodriguez-Gomez:2025hyy}. For more conventional superconducting materials with BCS behavior, $N=2$ and the effective Lagrangian does not have any fermion interaction higher than the usual $(\psi^\dagger\psi)^2$ interaction.
Higher fermion interactions emerge when the system involves $N\geq 4$, in particular, in multicomponent superconductors where fermions of different bands and/or orbitals contribute as different species \cite{2012RvMP...84.1383S, cao2018magic}.
In such materials, the impact of the  8-fermion coupling depends on its value at the scale of the cutoff. If it is sufficiently small, its contribution can be neglected. However, this coupling may play a role in systems with strong electron-phonon interaction such as type 1.5 superconductors.

In the following section, we will consider a different model where strong coupling, non-Fermi liquid effects naturally induce an 8-fermion vertex. This introduces new interesting features, as discussed below.

\section{Non-Fermi liquid fixed points in presence of $(\psi^\dagger \psi)^4$}\label{sec: NFL}

The aim of this section is to analyse the behaviour of the 8-fermion interaction within a quantum critical metallic system governed by non-Fermi liquid (NFL) dynamics. The latter has been studied in $d=2$ in \cite{Damia:2019bdx,Damia:2020yiu,Damia:2020bur}, based on ideas discussed in previous works \cite{Raghu:2015sna,Wang:2016hir,Fitzpatrick:2014cfa}. We begin by briefly reviewing the quantum critical point, driven by the inclusion of a non-relativistic critical boson transforming in the adjoint of $U(N)$ and interacting with the FS via a Yukawa coupling. In momentum space, the scalar-fermion system is described by the following Lagrangian
\begin{align}
L=&  \frac12\phi_i^j\left(q^2 + M_D^2\frac{|\Omega|}{q}\right)\phi_j^i + \psi_i^{\dagger}\left(i \omega -v_0 k_\perp \right)\psi^i \nonumber\\
& + \frac{g_0}{\sqrt{N}} \phi^i_j(\Omega,\vec q) \psi_i^\dagger(\omega,\vec p)\psi^j(\omega-\Omega,\vec p-\vec q)\ ,
\end{align}
where the mass of the boson has been tuned to zero to approach criticality. Notice that the dispersion relation considered above imposes for the bosonic frequency to scale as $\Omega\sim q^{z_b}$ with dynamical exponent $z_b=3$. The Debye mass $M_D\ll k_F$ determines the scale below which the $z_b=3$ scaling dominates over the usual relativistic one, hence playing the role of a UV cutoff for the effective field theory.\footnote{As shown in \cite{Damia:2019bdx}, if one
starts from a relativistic critical boson then the $z_b=3$ scaling is naturally induced by fermionic loops. In the context of the present work, we will not commit to any particular UV completion.} The reason behind the choice of this model is that it flows to a quantum critical point with non-Fermi liquid dynamics \cite{Damia:2019bdx}, as we briefly review below. The interactions are mediated by the (bare) relevant coupling $g_0$ of mass dimension $[g_0]=1/2$.

In the large $N$ limit we can neglect the quantum corrections to the Yukawa vertex. The strong coupling dynamics is thus reflected in a large anomalous dimension acquired by the fermions, encoded in the quantum self-energy
\be
\label{selfenergy}
\Sigma(\omega)= \Lambda_{\rm NFL}^{1/3} {\rm sgn}(\omega)|\omega|^{2/3} \,\, , 
\ee
with
\be
\Lambda_{\rm NFL}=\frac{g_0^6}{(2\pi v_0 \sqrt{3} )^3 M_D^2}\ ,
\ee
driving the system to a non-trivial fixed point with fermionic dynamical exponent $z_f=2/3$ at energy scales below $\Lambda_{\rm NFL}$. This picture has been assessed non-perturbatively by finding a self-consistent solution to the Schwinger-Dyson-Eliashberg equations both at zero and finite temperature \cite{Damia:2019bdx,Damia:2020bur}.

In this work, we restrict our analysis to the perturbative description of the fixed point. Quantum corrections naturally organize in powers of the following combination
\be\label{eq: alpha}
\alpha_0=\frac{g_0^2}{6\pi \sqrt{3} v_0}\ .
\ee
The renormalized fermionic field is defined by $\psi = Z^{-1/2}\psi_r$. Using \eqref{selfenergy}
we find
\be
Z(\mu)=1+3(M_D^2\mu)^{-1/3}\alpha_0\ .
\ee
The dimensionless running coupling $\alpha(\mu)$ is then given by
\be
\alpha(\mu)= (M_D^2\mu)^{-1/3} Z(\mu)^{-1}\alpha_0\ .
\ee
The renormalization procedure leads to a renormalized Fermi velocity
\be
v(\mu)=Z(\mu)^{-1}v_0\ .
\ee

The running of the renormalized Yukawa coupling is governed by the beta function
\be\label{eq: NFL beta alfa}
\mu\partial_\mu \alpha= -\frac{\alpha}{3} + 2\gamma\alpha= -\frac{\alpha}{3} + \alpha^2\ ,
\ee
where we used that $2\gamma=-\mu\partial_\mu\log Z=\alpha$. The quantum critical point originates as a solution to the above equation, where the coupling $\alpha$ flows to a finite value
\be
\label{aster}
\alpha^*=\frac13\ .
\ee

Notice that the Fermi velocity vanishes as $(\mu/\Lambda_{\rm NFL})^{1/3}$ at the fixed point, signalling the breakdown of the Fermi liquid description in terms of Landau quasiparticles. The damped critical boson also leads to a reorganization of the perturbative expansion in powers of the coupling \eqref{eq: alpha}, as opposed to $g_0^2$, and we refer the reader to \cite{Fitzpatrick:2014cfa} for an extensive analysis of this and related phenomena. 

The emergence of NFL dynamics makes it more convenient to parametrize the fermion couplings by including powers of the Fermi velocity \cite{Fitzpatrick:2014cfa}. A natural choice is the following
\be\label{eq: normalizations}
g_4=\frac{v_0 \lambda_{0}}{k_F} \quad , \quad g_8=\frac{v_0^3 \kappa_{0}}{k_F^3}\ ,
\ee
where we have  included the corresponding powers of the Fermi momentum in order to match the tree-level scaling dimensions.

\begin{figure}[h!]
\centering
\includegraphics[scale=.25]{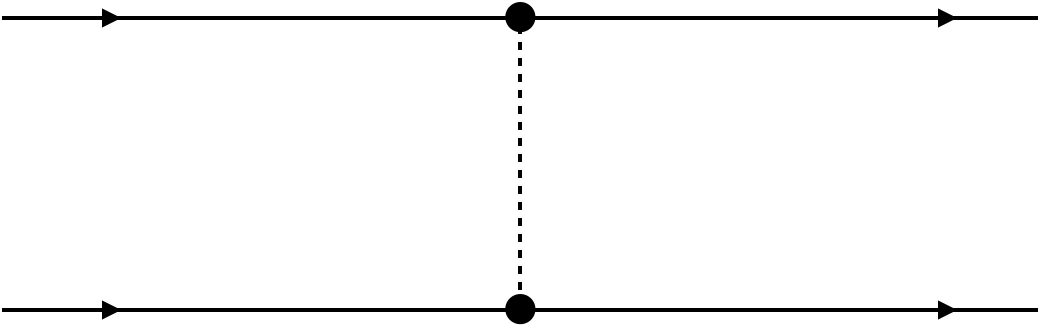}
\caption{Boson exchange contribution to the BCS channel.}
\label{4FermiInduced}
\end{figure}

As originally noticed in the context of colour superconductivity \cite{Son:1998uk}, the boson-fermion dynamics also induces a {\it relevant} 4-fermion coupling in the BCS channel. This effect is produced by boson exchange described by the diagram  in Fig.~\ref{4FermiInduced} and gives rise to the following contribution
\be
-\frac{4\pi v_0}{k_F N}\frac{\alpha_0}{\left(M_D^2\mu \right)^{1/3}}\ ,
\ee
where $\mu$ is identified with the external frequency carried by the boson. This singular contribution originates from the special RG flow taking place in the boson-fermion system, allowing for the mediation of high-energy bosonic modes within low-energy fermionic processes. This type of UV/IR mixing effects have been thoroughly studied in \cite{Son:1998uk,Fitzpatrick:2014cfa,Metlitski}.

\begin{figure*}[t]
    \centering
\includegraphics[width=\textwidth,height=0.2\textheight,keepaspectratio
]{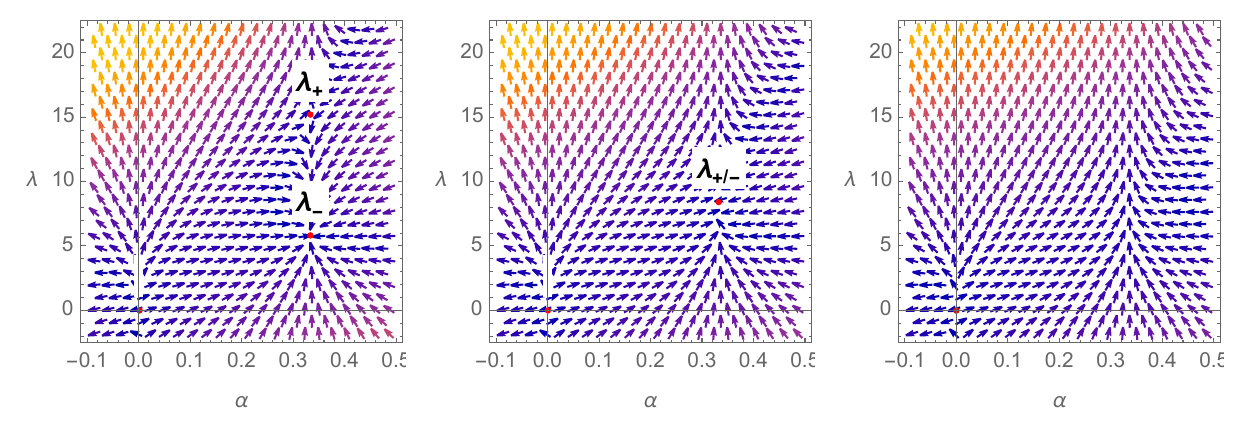}
    \caption{RG-flow trajectories in terms of the RG flow time $t=-\log \mu$ (which grows towards the IR) in the $(\alpha,\lambda)$ plane for $N=10,8,6$.}
    \label{Annihilation}
\end{figure*}

Collecting all the contributions, the effective 4-fermion  coupling can be extracted from the correlation function of renormalized fermion fields at vanishing external momenta
\be
\lambda(\mu)=Z(\mu)^{-1}\left[\lambda_{0}+ \frac{4\pi\alpha_0}{\left(M_D^2\mu\right)^{1/3}} +\frac{\lambda_{0}^2}{2\pi N}\log\frac{\Lambda}{\mu} \right]\ ,
\ee
thus leading to the beta function
\be\label{eq: NFL 4-fermi beta}
\mu\partial_\mu\lambda = -\frac{4\pi}{3}\alpha +\alpha \lambda -\frac{1}{2\pi N}\lambda^2\ .
\ee
Therefore, to this order, the beta function contains three contributions: the boson exchange,
the fermion anomalous dimension and the one-loop contribution \eqref{eq: fermion bubble}.

The solution of the fixed point equations $\beta_\alpha=\beta_\lambda=0$ gives $\alpha^*=\frac13$ and \cite{Raghu:2015sna}
\be\label{eq: lambda fixed points}
\lambda_{\pm}=\frac{\pi N}{3}\left(1\pm \eta\right)\ ,\quad \eta\equiv \sqrt{1-\frac{8}{N}}\ .
\ee
The solutions are real only for $N\geq N_{\rm cr}$, with $N_{\rm cr}=8$.
When $N=N_{\rm cr}$, $\lambda_+=\lambda_-=\frac{8\pi}{3}$ is an unstable UV fixed point.
When $N>N_{\rm cr}$,
$\lambda^*\equiv\lambda_-$ is a stable
 IR fixed point. As $N$ is decreased towards $N_{\rm cr}$, the two fixed points approach and annihilate for $N\to N_{\rm cr}$, where the system develops a non-analytic BKT scaling as a function of $N-N_{\rm cr}$ \cite{Damia:2020bur}, similar to the mechanism described in \cite{Son:1998uk}.
In turn, for $N<N_{\rm cr}$, the 4-fermi interaction grows unbounded signalling the onset of a superconducting instability. We refer to \cite{Raghu:2015sna,Damia:2019bdx,Damia:2020bur} for previous studies of this type of fixed points in various scenarios, both at zero and finite temperature. The RG trajectories in Fig.~\ref{Annihilation} illustrate this behaviour.

Comparing \eqref{eq: NFL 4-fermi beta} to \eqref{eq: 4-fermi beta}, one notices that the strongly coupled boson-fermion system leads to two competing effects. On the one hand, exchange of high energy bosons tends to increase the tendency to a pairing instability, by means of a relevant tree-level contribution. On the top of that, the large anomalous dimension acquired by the fermions, encoded in the wave function renormalization $Z(\mu)$, tends to destroy the quasiparticles and subsequently suppresses the formation of Cooper pairs. From \eqref{eq: lambda fixed points} one concludes that this interplay leads to fixed points
described by a non-Fermi liquid with {\it finite} BCS coupling,
as long as $N>N_{\rm cr}$.

\begin{figure*}[t]
    \centering
\includegraphics[width=\textwidth,height=0.2\textheight,keepaspectratio
]{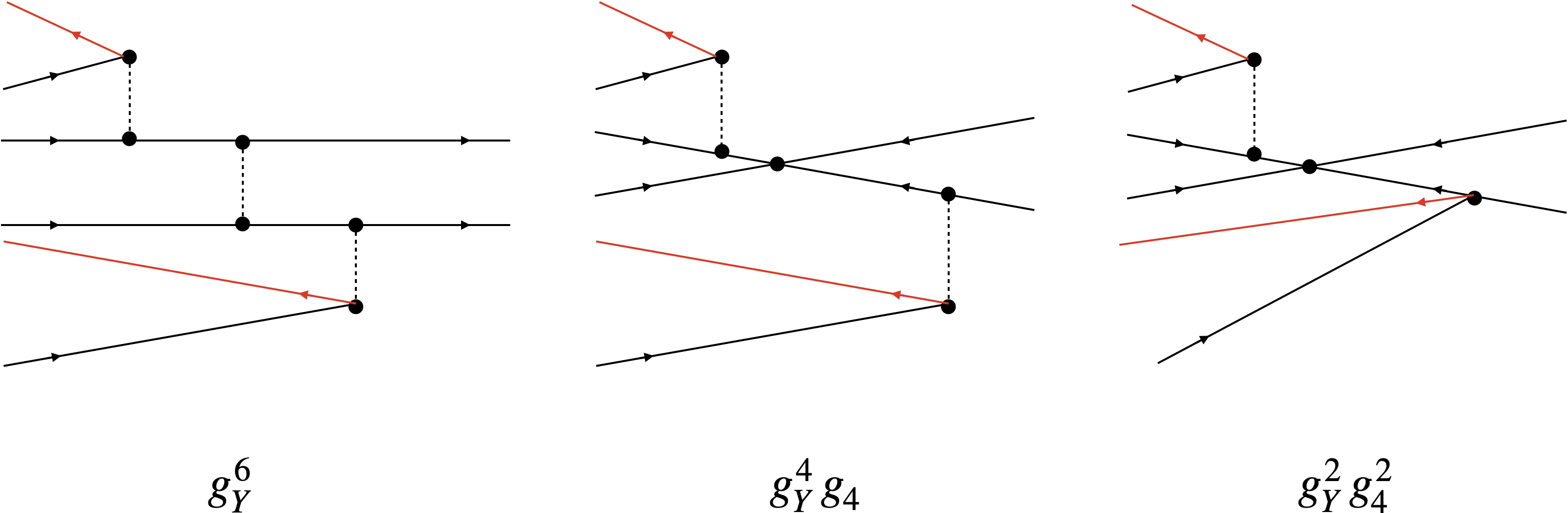}
    \caption{A sample of tree-level 8-fermi diagrams involving Yukawa interactions.}
    \label{8FermiInduced}
\end{figure*}

We now proceed to include the 8-fermion interaction in the channel comprised by \eqref{eq: 8 fermi BCS}. In addition to the contribution of the diagram on panel $b)$ of Fig.~\ref{Diagrams},
 there are  contributions originating from the interactions with the critical bosons. One of these is the anomalous dimension induced by the wave function renormalization $Z(\mu)$. In addition, there is a large set of diagrams contributing to the boson exchange channel, with some representatives shown in Fig.~\ref{8FermiInduced}.  Because these contributions do not depend on the 8-fermion coupling, they will not affect the analysis of stability of fixed points. 
 
As a consequence of our choice \eqref{eq: normalizations} for the normalization of the vertex, the contributions of  the type of Fig.~\ref{8FermiInduced} are naturally comprised 
in a certain function $F(\alpha_0,\lambda_{0})$ of the remaining couplings. 
This has the form
$$
F(\alpha_0,\lambda_{0})=m_1 \mu^{-1} \alpha_0^3 + m_2\mu^{-\frac23}\alpha_0^2\lambda_0+m_3\mu^{-\frac13}\alpha_0\lambda_0^2\ ,
$$
where $m_i$ are dimensionless coefficients that are independent of the couplings. Keeping terms up to quadratic order in $\kappa_0$ and introducing the dimensionless running coupling $\kappa=\mu^{\Delta_8}\kappa_0$, one obtains  
\be
\kappa(\mu)= Z(\mu)^{-1}\left[\mu^{\Delta_8}\kappa_{0}\left(1+\frac{\lambda_{0}}{2\pi N}\log\frac{\Lambda}{\mu}\right)+ F(\alpha_0,\lambda_{0})  \right]
\ee
The effective running coupling is therefore governed by
\be\label{eq: NFL 8-fermi beta}
\mu\partial_\mu \kappa = - f(\alpha,\lambda) +(\Delta_8+\alpha)\kappa -\frac{\lambda\kappa}{2\pi N}\ ,
\ee
with $f\equiv - Z^{-1}\mu\partial_\mu F$. It is worth stressing that, even if one starts with $\kappa=0$, a non-zero $\kappa $ is induced by the RG flow.
In fact, in the  present model, boson-exchange Feynman diagrams induce the infinite tower of fermionic operators $(\psi^\dagger \psi)^{2n}$. A consistent and complete RG analysis therefore requires the inclusion of all such interactions.

The renormalization group flow of the system is thus accounted for the system of equations \eqref{eq: NFL beta alfa}, \eqref{eq: NFL 4-fermi beta} and \eqref{eq: NFL 8-fermi beta}. Importantly, these equations have a sequential structure: \eqref{eq: NFL beta alfa} involves  only $\alpha$ dependence; \eqref{eq: NFL 4-fermi beta} involves $\alpha$ and $\lambda$ dependence; \eqref{eq: NFL 8-fermi beta} involves $\alpha$, $\lambda$ and $\kappa$ dependence. Because of this structure, the RG trajectories in the $\{\alpha,\lambda\}$ plane are identical to those shown in Fig.~\ref{Annihilation}, and the  coupling $\kappa$ simply provides a transverse direction.

Let us examine the fixed-point structure of the
beta functions \eqref{eq: NFL beta alfa}, \eqref{eq: NFL 4-fermi beta} and \eqref{eq: NFL 8-fermi beta}.
We first consider $N\geq N_{\rm cr}$, with $N_{\rm cr}=8$.
In  $(\alpha,\lambda,\kappa)$ space, there are three fixed points, located at 
$$
(0,0,0)\ , \quad (\alpha^*,\lambda_+,\kappa_+) \ ,\quad (\alpha^*,\lambda_-,\kappa_-)\ , 
$$
with $\alpha^*, \ \lambda_\pm$ given by \eqref{aster}, \eqref{eq: lambda fixed points} and
\be\label{eq: NFL 8 fixed}
\kappa_\pm= \frac{6 f(\alpha^*,\lambda_\pm)}{13\mp \sqrt{1-\frac{8}{N}}}\ .
\ee

We now consider the stability of the fixed points.
The RG trajectories in the $\{\alpha,\lambda\}$ plane  shown in Fig.~\ref{Annihilation} indicate that $(0,0)$ and $(\alpha^*,\lambda_+)$ are unstable, and,  at least for the flow in this plane, $(\alpha^*,\lambda_-)$
is stable. This naturally raises the question of whether the new eight-fermion interaction 
 can destabilize the fixed point  $(\alpha^*,\lambda_-,\kappa_-)$ when the RG trajectories in the full three‑dimensional coupling space are taken into account.

To study  stability, we linearize the beta functions around  each fixed point. In particular, for $(\alpha^*,\lambda_-,\kappa_-)$ we write $\alpha=\alpha^{\star}+\delta \alpha$, $\lambda=\lambda_-+\delta\lambda $ and $\kappa=\kappa_-+\delta\kappa$. In terms of $\vec{\delta}=(\delta\alpha,\delta\lambda,\delta\kappa)$, we have an equation $\partial_t\vec{\delta}=\mathbb{V}\,\vec{\delta}$ with $\mathbb{V}$ a $3\times 3$ matrix and $t$ is the RG flow time $t=-\log 
\mu$ which grows towards the IR. Due to the sequential structure of the beta functions, not only the eigenvalues along $\{\alpha,\lambda\}$ plane are identical to the case without $\kappa$, but also, importantly, the matrix $\mathbb{V}$ does not depend on $f(\alpha,\lambda)$. Computing its eigenvalues, we find 
 \begin{equation}
\begin{array}{lcl}(\alpha^*,\lambda_{\pm},\kappa_{\pm}) & \leadsto& \left\{   -\frac13,\pm \frac{\eta}{3},\ -\frac16 (1+\eta)  \right\}\ ,\\ \\
 (0,0,0)&\leadsto&\left\{  0,\frac{1}{6},1 \right\}\ .
 \end{array}
\end{equation}
Thus we see that only for the fixed point $(\alpha^*,\lambda_-,\kappa_-)$ all eigenvalues are negative for $N>8$,
which shows that the IR fixed point remains stable.
 The system describes a metallic fixed point with non-Fermi liquid dynamics and finite BCS coupling.
As $N$ decreases towards $N_{\rm cr}$, $(\alpha^*,\lambda_{\pm},\kappa_{\pm})$ approach (and eventually annihilate) at $\lambda_0$ as in the middle panel of Fig.~\ref{Annihilation} at a value of $\kappa=\frac{6f(\alpha^*,\lambda_0)}{13}$.

\medskip

Finally, consider  $N< N_{\rm cr}$. In this case there is a single fixed point located at the origin in the $\{\alpha,\lambda,\kappa\}$ space. This point is unstable, with  the couplings $\lambda$ and $\kappa$ flowing to infinity in the IR. Thus, the system is driven to a superconducting phase due to the divergent 4-fermion coupling.

\medskip

An interesting direction for future work is to explore the theory in the presence of defects, which play the role of impurities. In particular, it would be interesting to study the non-Fermi-liquid  behaviour at the fixed point $(\alpha^*,\lambda_-,\kappa_-)$  as the defect parameter is varied, and to determine whether the system exhibits quantum critical points at critical values of this parameter. It would  be instructive to investigate the effect of a non-zero coupling $\kappa_-$  on the free energy and the associated critical behaviour.

\section{Applications}\label{sec:applications}

For systems with 
$N\geq 4$
 fermionic species, higher‑order interactions naturally appear, particularly in multicomponent superconductors where different bands contribute separate fermion species.
These fermionic interactions may play an important role in systems with strong electron--phonon coupling.
In particular, it was shown in \cite{Rodriguez-Gomez:2025hyy} that the presence of a $ \psi^8 $ interaction modifies the temperature dependence of the energy gap, yielding a flatter behaviour over most of the temperature range and a more rapid decrease near the critical temperature. If the $g_8$ coupling is  increased further, the transition becomes first order once 
$g_8$ exceeds a critical value.
The free energy develops qualitatively new features, in particular, differences in the temperature dependence of the specific heat in the interval $ T \in (0, T_c) $.

In conventional BCS theory, the zero-temperature energy gap  $\Delta_{\rm gap} \equiv \Delta(0) $ scales linearly with the critical temperature. 
The critical temperature is
\be
\label{critT}
T_c\approx b M_D\, e^{-1/g_4v_0}\ ,
\ee
where $b$ is a numerical constant of order 1.

The addition of $g_{2n}$ interactions does not modify this formula as long as the higher couplings $g_{4n}$ (with $n>1$) are small enough to preserve the second-order nature of the transition. At $T_c$, the $\psi\psi$ condensate goes to zero, so higher-order fermion interactions can be  neglected.

On the other hand $\Delta_{\rm gap}$ is generally affected by
higher-fermion interactions, as $\Delta_{\rm gap}$  does not need to be small. 
In standard BCS theory, one  finds
\be
 \Delta_{\rm gap} =\frac{M_D}{\sinh(1/g_4v_0)}\approx 2M_D e^{-1/g_4v_0}\ .
\ee
This gives
\be
T_c\sim \Delta_{\rm gap} \ .
\ee
By contrast, in the presence of a $ \psi^8 $ interaction this relationship becomes nonlinear.
Solving the gap equation,  one finds \cite{Rodriguez-Gomez:2025hyy}
\be
\label{gap0}
\eta_0\, \Delta_{\rm gap} =\frac{M_D}{\sinh\big[1/(g_4 v_0 \eta_0)\big]}\ ,\qquad 
 \ee
 where
\be
\eta_0 =1+\frac{g_8}{g_4^3}\, \Delta_{\rm gap}^2\ .
\ee
Combining \eqref{gap0} with \eqref{critT}, one finds that the ratio $ \Delta_{\rm gap} / T_c $,  exhibits an interesting nonlinear behaviour.
 For sufficiently small $g_8$, one finds
 \be
T_c
\sim \Delta_{\rm gap} \ e^{-g_8\Delta_{\rm gap}^2/(g_4^4v_0)}\ .
\label{gapdelta}
\ee

Higher-order fermionic couplings may also have important implications for metallic systems governed by non-Fermi-liquid dynamics. Typically one finds scale separation between the critical temperature and the gap, namely $\Delta_{\rm gap}\sim \big(T_c\big)^\gamma$ with $\gamma<1$. For the model considered in section \ref{sec: NFL}, one finds $\gamma=2/3$ \cite{Damia:2020bur} (hierarchies among these quantities are relevant in experimental studies in non-conventional superconductors \cite{2012JPSJ...81a1006Y}). 
A non-perturbative analysis along the lines of \cite{Damia:2020bur} may enable to understand how this scaling gets modified by the higher order interactions.

\medskip

The ratio $\Delta_{\rm gap}/T_c$ can be measured experimentally with high precision. A power-law relation of the form
$
\Delta_{\rm gap}\sim \left(T_c\right)^\gamma$
gives rise to  a linear dependence in a plot of $\log(\Delta_{\rm gap}/T_c)$ versus $\Delta$. 
By contrast, Eq. \eqref{gapdelta} implies that the presence of a nonzero $g_8$ coupling leads to markedly distinct behaviour, namely,
$$
\log\left( \frac{\Delta_{\rm gap}}{T_c}\right)\sim c_0+\frac{g_8}{g_4^4v_0}\Delta_{\rm gap}^2\ , 
$$
where $c_0$ is a constant. This prediction could be tested experimentally by investigating whether such nonlinear signatures can be detected in real materials. 
In particular, materials such as those studied in \cite{khasanov2018superconductivity}, whose energy-gap temperature dependence exhibits signatures of strong electron--phonon coupling, may provide a promising testing ground for these effects.

\medskip

\makeatletter
\begingroup
\let\addcontentsline\@gobblethree
\section*{Acknowledgements}
JGR and JAD acknowledge financial support from the Spanish  MCIN/AEI/10.13039/501100011033 grant PID2022-126224NB-C21. D.R.-G is supported in part by the Spanish national grant MCIU-22-PID2021-123021NB-I00.
\endgroup
\makeatother

\bibliographystyle{apsrev4-2}
\bibliography{NFL}

@article{khasanov2018superconductivity,
  title={Superconductivity of Bi-III phase of elemental bismuth: Insights from muon-spin rotation and density functional theory},
  author={Khasanov, Rustem and Luetkens, Hubertus and Morenzoni, Elvezio and Simutis, Gediminas and Sch{\"o}necker, Stephan and {\"O}stlin, Andreas and Chioncel, Liviu and Amato, Alex},
  journal={Physical Review B},
  volume={98},
  number={14},
  pages={140504},
  year={2018},
  publisher={APS}
}

@article{Kuo2016,
  title={Ubiquitous signatures of nematic quantum criticality in optimally doped Fe-based superconductors},
  author={Kuo, H. and Chu, J. and Palmstrom, J. and Kivelson, S. and Fisher, I.},
  journal={Science},
  volume={352},
  number={6288},
  pages={958--962},
  year={2016},
  publisher={American Association for the Advancement of Science}
}

@article{Schofield1999,
  title={Non-Fermi liquids},
  author={Schofield, A.},
  journal={Contemporary Physics},
  volume={40},
  number={2},
  pages={95--115},
  year={1999},
  publisher={Taylor \& Francis}
}

@article{Shibauchi2013,
  title={Quantum critical point lying beneath the superconducting dome in iron-pnictides},
  author={Shibauchi, T. and Carrington, A. and Matsuda, Y.},
  journal={arXiv preprint arXiv:1304.6387},
  year={2013}
}

@article{Shankar,
      author         = "Shankar, R.",
      title          = "{Renormalization group approach to interacting fermions}",
      journal        = "Rev.Mod.Phys.",
      volume         = "66",
      pages          = "129-192",
      doi            = "10.1103/RevModPhys.66.129",
      year           = "1994",
      reportNumber   = "PRINT-93-0200-REV (YALE), PRINT-93-0200 (YALE)",
      SLACcitation   = "%%CITATION = RMPHA,66,129;%%",
}

@article{Lee2009,
	Author = {Lee, S-S.},
	Doi = {10.1103/PhysRevB.80.165102},
	Issue = {16},
	Journal = {Phys. Rev. B},
	Month = {Oct},
	Numpages = {13},
	Pages = {165102},
	Publisher = {American Physical Society},
	Title = {Low-energy effective theory of Fermi surface coupled with U(1) gauge field in $2+1$ dimensions},
	Volume = {80},
	Year = {2009}}

@article{Altshuler1994,
	Author = {Altshuler, B. L. and Ioffe, L. B. and Millis, A. J.},
	Doi = {10.1103/PhysRevB.50.14048},
	Issue = {19},
	Journal = {Phys. Rev. B},
	Month = {Nov},
	Pages = {14048--14064},
	Publisher = {American Physical Society},
	Title = {Low-energy properties of fermions with singular interactions},
	Volume = {50},
	Year = {1994}}

@misc{Polchinski,
  author = {Polchinski, J.},
  title = {Effective field theory and the Fermi surface},
  year = {1992},
  eprint = {hep-th/9210046},
  archivePrefix = {arXiv},
  primaryClass = {hep-th}
}

@article{Hertz1976,
	Author = {Hertz, J.},
	Doi = {10.1103/PhysRevB.14.1165},
	Issue = {3},
	Journal = {Phys. Rev. B},
	Month = {Aug},
	Pages = {1165--1184},
	Publisher = {American Physical Society},
	Title = {Quantum critical phenomena},
	Volume = {14},
	Year = {1976}}

@article{Metlitski2010,
	Author = {Metlitski, M. and Sachdev, S.},
	Doi = {10.1103/PhysRevB.82.075127},
	Issue = {7},
	Journal = {Phys. Rev. B},
	Month = {Aug},
	Numpages = {24},
	Pages = {075127},
	Publisher = {American Physical Society},
	Title = {Quantum phase transitions of metals in two spatial dimensions. I. Ising-nematic order},
	Volume = {82},
	Year = {2010}}

@article{Nayak1994,
	Author = {Nayak, C. and Wilczek, F.},
	Journal = {Nuclear Physics B},
	Number = {3},
	Pages = {359--373},
	Publisher = {Elsevier},
	Title = {Non-Fermi liquid fixed point in 2+ 1 dimensions},
	Volume = {417},
	Year = {1994}}

@article{Mross2010,
	Author = {Mross, D. and McGreevy, J. and Liu, H. and Senthil, T.},
	Doi = {10.1103/PhysRevB.82.045121},
	Issue = {4},
	Journal = {Phys. Rev. B},
	Month = {Jul},
	Numpages = {19},
	Pages = {045121},
	Publisher = {American Physical Society},
	Title = {Controlled expansion for certain non-Fermi-liquid metals},
	Volume = {82},
	Year = {2010}}

@article{Metlitski,
  title = {Cooper pairing in non-Fermi liquids},
  author = {Metlitski, M. and Mross, D. and Sachdev, S. and Senthil, T.},
  journal = {Phys. Rev. B},
  volume = {91},
  issue = {11},
  pages = {115111},
  numpages = {18},
  year = {2015},
  month = {Mar},
  publisher = {American Physical Society},
  doi = {10.1103/PhysRevB.91.115111},
  url = {http://link.aps.org/doi/10.1103/PhysRevB.91.115111}
}

@article{Fitzpatrick:2014cfa,
      author         = "Fitzpatrick, A. and Torroba, G. and Wang,
                        H.",
      title          = "{Aspects of Renormalization in Finite Density Field
                        Theory}",
      journal        = "Phys.Rev.",
      volume         = "B91",
      pages          = "195135",
      doi            = "10.1103/PhysRevB.91.195135",
      year           = "2015",
      eprint         = "1410.6811",
      archivePrefix  = "arXiv",
      primaryClass   = "cond-mat.str-el",
      SLACcitation   = "%%CITATION = ARXIV:1410.6811;%%",
}

@article{Son:1998uk,
  title = {Superconductivity by long-range color magnetic interaction in high-density quark matter},
  author = {Son, D. T.},
  journal = {Phys. Rev. D},
  volume = {59},
  issue = {9},
  pages = {094019},
  numpages = {8},
  year = {1999},
  month = {Apr},
  doi = {10.1103/PhysRevD.59.094019},
  publisher = {American Physical Society}
}

@article{Raghu:2015sna,
      author         = "Raghu, S. and Torroba, G. and Wang, H.",
      title          = "{Metallic quantum critical points with finite BCS
                        couplings}",
      journal        = "Phys. Rev.",
      volume         = "B92",
      year           = "2015",
      number         = "20",
      pages          = "205104",
      doi            = "10.1103/PhysRevB.92.205104",
      eprint         = "1507.06652",
      archivePrefix  = "arXiv",
      primaryClass   = "cond-mat.str-el",
      SLACcitation   = "%%CITATION = ARXIV:1507.06652;%%"
}

@article{Wang:2016hir,
      author         = "Wang, H. and Raghu, S. and Torroba, G.",
      title          = "{Non-Fermi liquid Superconductivity: Eliashberg versus
                        the Renormalization Group}",
      journal        = "Phys. Rev.",
      volume         = "B95",
      year           = "2017",
      number         = "16",
      pages          = "165137",
      doi            = "10.1103/PhysRevB.95.165137",
      eprint         = "1612.01971",
      archivePrefix  = "arXiv",
      primaryClass   = "cond-mat.str-el",
      SLACcitation   = "%%CITATION = ARXIV:1612.01971;%%"
}

@article{Damia:2019bdx,
      author         = "Aguilera Damia, Jeremias and Kachru, Shamit and Raghu,
                        Srinivas and Torroba, Gonzalo",
      title          = "{Two dimensional non-Fermi liquid metals: a solvable
                        large N limit}",
      journal        = "Phys. Rev. Lett.",
      volume         = "123",
      year           = "2019",
      number         = "9",
      pages          = "096402",
      doi            = "10.1103/PhysRevLett.123.096402",
      eprint         = "1905.08256",
      archivePrefix  = "arXiv",
      primaryClass   = "cond-mat.str-el",
      SLACcitation   = "%%CITATION = ARXIV:1905.08256;%%"
}

@article{Damia:2020yiu,
    author = "Damia, Jeremias Aguilera and Solis, Mario and Torroba, Gonzalo",
    title = "{How non-Fermi liquids cure their infrared divergences}",
    eprint = "2004.05181",
    archivePrefix = "arXiv",
    primaryClass = "cond-mat.str-el",
    doi = "10.1103/PhysRevB.102.045147",
    journal = "Phys. Rev. B",
    volume = "102",
    number = "4",
    pages = "045147",
    year = "2020"
}

@misc{Kivelson-Zaanen-review,
    title={High Temperature Superconductivity in the Cuprates},
    author={B. Keimer and S. A. Kivelson and M. R. Norman and S. Uchida and J. Zaanen},
    year={2014},
    eprint={1409.4673},
    archivePrefix={arXiv},
    primaryClass={cond-mat.supr-con}
}

@ARTICLE{2012RvMP...84.1383S,
       author = {{Scalapino}, D.~J.},
        title = "{A common thread: The pairing interaction for unconventional superconductors}",
      journal = {Reviews of Modern Physics},
         year = 2012,
        month = oct,
       volume = {84},
       number = {4},
        pages = {1383-1417},
          doi = {10.1103/RevModPhys.84.1383},
archivePrefix = {arXiv},
       eprint = {1207.4093},
 primaryClass = {cond-mat.supr-con},
       adsurl = {https://ui.adsabs.harvard.edu/abs/2012RvMP...84.1383S}
}

@article{cao2018magic,
  title={Magic-angle graphene superlattices: a new platform for unconventional superconductivity},
  author={Cao, Yuan and Fatemi, Valla and Fang, Shiang and Watanabe, Kenji and Taniguchi, Takashi and Kaxiras, Efthimios and Jarillo-Herrero, Pablo},
  journal={arXiv preprint arXiv:1803.02342},
  year={2018}
}

@ARTICLE{2012JPSJ...81a1006Y,
       author = {{Yoshida}, Teppei and {Hashimoto}, Makoto and {Vishik}, Inna M. and
         {Shen}, Zhi-Xun and {Fujimori}, Atsushi},
        title = "{Pseudogap, Superconducting Gap, and Fermi Arc in High-T$_{c}$ Cuprates Revealed by Angle-Resolved Photoemission Spectroscopy}",
      journal = {Journal of the Physical Society of Japan},
         year = 2012,
        month = jan,
       volume = {81},
       number = {1},
        pages = {011006-011006},
          doi = {10.1143/JPSJ.81.011006},
archivePrefix = {arXiv},
       eprint = {1203.0600},
 primaryClass = {cond-mat.supr-con},
       adsurl = {https://ui.adsabs.harvard.edu/abs/2012JPSJ...81a1006Y}
}

@article{Damia:2020bur,
    author = "Damia, Jeremias Aguilera and Sol{\'\i}s, Mario and Torroba, Gonzalo",
    title = "{Thermal effects in non-Fermi liquid superconductivity}",
    eprint = "2009.11887",
    archivePrefix = "arXiv",
    primaryClass = "cond-mat.str-el",
    doi = "10.1103/PhysRevB.103.155161",
    journal = "Phys. Rev. B",
    volume = "103",
    number = "15",
    pages = "155161",
    year = "2021"
}

@article{Rodriguez-Gomez:2025hyy,
    author = "Rodriguez-Gomez, D. and Russo, J. G.",
    title = "{Higher-Order Fermion Interactions in Effective Field Theories for Phase Transitions}",
    eprint = "2507.11624",
    archivePrefix = "arXiv",
    primaryClass = "cond-mat.supr-con",
    doi = "10.1016/j.physletb.2026.140172",
    journal = "Phys. Lett. B",
    volume = "873",
    pages = "140172",
    year = "2026"
}

\end{document}